\documentclass{article}
\usepackage[dvips]{graphicx}
\usepackage{cite}
\begin{document}
\title{Recognizing number of communities and detecting community structures in complex networks}
\author{\textbf{Hongjue Wang}\\School of Systems Science\\Beijing Normal University\\Beijing, 100875, China
\\\textbf{Tao Wang}\\School of Mathematics and Physics\\North China Electric Power University\\Baoding, HeBei, 071003,China}

\date{}\maketitle
\begin{abstract}
Recognizing number of communities and detecting community structures of complex network are discussed in this paper. As a visual and feasible algorithm, block model has been successfully applied to detect community structures in complex network. In order to measure the quality of the block model, we first define an objective function $WQ$ value. For obtaining block model $B$ of a network, GSA algorithm is applied to optimize $WQ$ with the help of random keys. After executing processes $AO$ (Adding Ones) and $RO$ (Removing Ones) on block model $B$, the number of communities of a network can be recognized distinctly. Furthermore, based on the advantage of block model that its sort order of nodes is in correspondence with sort order of communities, so a new fuzzy boundary algorithm for detecting community structures is proposed and successfully applied to some representative networks. Finally, experimental results demonstrate the feasibility of the proposed algorithm.

{\bf Keywords: }Community structures, Block model, Complex network, GSA algorithm
\end{abstract}

\section{Introduction}
In recent years, detecting community structure in complex networks has attracted a growing attention of many researchers ~\cite{1wang2015novel,2faqeeh2012community}. Community structures widely exist in many networked systems including friendship networks~\cite{1wang2015novel}, coauthor networks, protein-protein interaction network ~\cite{3okamura2015comparison,4fortunato2010community}, mobile phone networks~\cite{5cucuringu2013extracting}, WWW networks~\cite{6adamic2005political}, transportation networks~\cite{7roth2012long}, online shopping networks, animal relationship networks, etc. However, it is very difficult to provide an accurate and recognized definition for community structure, due to its idiosyncratic middle scale property and the complexity and heterogeneity of these real networks in our society~\cite{8li2013efficient}. It is generally thought that community structures are subgroups of a network with dense internal links and sparse external links. As one of the most important tasks in complex networks researches, detecting community structure can be utilized for researching network topological structure, determining cause of group formation~\cite{9tang2012community}, understanding network underlying laws ~\cite{2faqeeh2012community,10cong2014approaching}, learning network functions~\cite{11huang2013decentralized,12he2015fast} and discovering group evolution~\cite{12he2015fast}, etc.

A great deal of novel approaches and technologies have been proposed successfully to discover communities in complex networks. Among these existing methods, after the measurement of algorithms modularity was introduced by Newman and Girvan ~\cite{13newman2004finding}，modularity optimization methods emerged with a large number in a brief period ~\cite{1wang2015novel,4fortunato2010community,14newman2004fast,15cafieri2014improving}. These methods include greedy optimization~\cite{16blondel2008fast}，hierarchical clustering~\cite{17queyroi2014optimizing}, extremal optimization~\cite{18suciu2015mixing}, spectral algorithm~\cite{19fu2010spectral,20richardson2009spectral}, intelligent algorithm ~\cite{21liu2007effective,22li2015overlapping,23hassan2015community}, mathematical programming~\cite{24xu2007finding,25bennett2012mathematical,26xu2010module} and so on. However, modularity optimization methods are NP-complete problem and modularity has the resolution limit problem due to its definition and selection of null model. So many researchers improved and extended modularity through great efforts, and some new algorithm measurements were also proposed such as partition density ~\cite{27ahn2010link}, normalized mutual information (NMI)~\cite{28chen2013detecting}, modularity density~\cite{29naeni2015ma}, local modularity~\cite{30xu2013parallel}, linked matrix factorization~\cite{31tang2009clustering}. For some other algorithms, such as core nodes based methods~\cite{28chen2013detecting,32wang2013detecting,33xin2013complex}, $k$-means methods ~\cite{4fortunato2010community,34liu2010detecting,35wang2012improved,36jinglu2013finding}，$k$-rank methods~\cite{38jiang2013efficient}, hierarchical clustering methods~\cite{4fortunato2010community} and spectral clustering~\cite{2faqeeh2012community} for more communities, they can obtain satisfactory community structures, but they are bound by the priori knowledge of the number of communities or a related threshold value~\cite{1wang2015novel,9tang2012community,11huang2013decentralized,33xin2013complex,38jiang2013efficient}. Identifying gaps between the eigenvalues of Laplacian matrix~\cite{4fortunato2010community} of a network or Clumpiness matrix~\cite{2faqeeh2012community} was often applied to estimate the number of communities. For networks with indistinct community boundaries and mixed communities, there are not distinct gaps between the eigenvalues of their Laplacian matrixes or Clumpiness matrixes. In order to detect communities precisely, Wei Tang et al. proposed a block model approximation algorithm called Linked Matrix Factorization Algorithm ~\cite{31tang2009clustering}. This algorithm needs a block indicator matrix $S$ that is costly obtained by calculating top $k$ eigenvectors of network adjacency matrix with maximum eigenvalues.

Considering that complex calculation of eigenvectors of matrices exists in the current block model approximation algorithms, a novel measurement WQ of block model is proposed according to the property that most elements with a value of one are around the leading diagonal in block model. Furthermore, GSA~\cite{39rashedi2009gsa,40xu2014improved} and Random Keys algorithms~\cite{41ceberio2012review} are employed to optimize objective function WQ value in order to obtain the block model of complex network. For some algorithms are bound by the number of communities, a method to recognize number of communities is proposed using strong block of the network. The proposed processes AO and RO are executed on block model and a strong block is obtained, then the number of communities can be counted artificially from strong block model. In order to detect communities in complex networks, a novel fuzzy boundary algorithm is proposed according to the advantage that the order of rows and columns of block model B corresponds with the order of communities. Experimental results demonstrate the feasibility of the proposed algorithm.

\section{Preliminaries and definitions}
A connected network with no node strength, link weight and direction is denoted as $G = (V,E)$, where $V$ and $E$ are the set of nodes and links, respectively. $\left| V \right| = n$ is the number of nodes and $\left| E \right| = m$ is the number of links of network $G$. The network $G$ can also be represented by its adjacency matrix $A$, in which elements represent connections of network nodes. When a link between nodes ${v_i}$ and ${v_j}$  exists, then ${a_{i,j}}=1$ and ${a_{i,j}}=0$ otherwise. The degree of node ${v_i}$ is denoted as ${k_i}$ and ${k_i} = \sum\limits_i {{a_{i,j}}}$. Let ${G_i}({V_i},{E_i})$ represents a sub-network of a partition of $G$, i.e., a community of network $G$.

\subsection{Definition}
\textbf{Definition 1.} $WQ$ value. We take adjacency matrix $A$ of network $G$ as a table $\Delta$ with ${n^2}$ cells and the size of each cell is 1. Element ${a_{i,j}}$ is a cell of $\Delta$. If  ${a_{i,j}}{\rm{ = }}1$, the element is called as 1-element. Then the distance between cell ${a_{i,j}}$ and leading diagonal of matrix $A$ is $\frac{{\sqrt 2 \left( {\left| {i - j} \right| + 1} \right)}}{2}$. So the $WQ$ value of network $G$ is defined as the sum of all distances from 1-elements to main diagonal divided by $2\sqrt 2 n(2n - 1)$.
\begin{small}
\begin{equation}
WQ = \frac{1}{{4n(2n - 1)}}\sum\limits_{1 \le i,j \le n} {\left( {\left| {i - j} \right| + 1} \right)}
\label{eq.1}
\end{equation}
\end{small}
Clearly, $WQ \in \left( {0,1} \right)$. In the following we will verify that less $WQ$ value corresponds more distinct blocks and two isomorphic networks may have different $WQ$ values.

\noindent
\textbf{Definition 2.} Airtightness Value. Suppose the cell of element ${a_{i,j}}$ has $r$ neighbor cells, then the airtightness value of element in ${a_{i,j}}$ is defined as

\begin{small}
\begin{equation}
H_{i,j}^A = \frac{{\sum\limits_{i-1 \le k \le i + 1,j - 1 \le l\le j+1}{{a_{k,l}}} }}{{r{\rm{ + }}1}}
\label{eq.2}
\end{equation}
\end{small}
\noindent
Clearly, $H_{i,j}^A \in \left( {0,1} \right)$ and the larger the airtightness value, the tighter the network structure.

\noindent
\textbf{Definition 3.} Block. For a network $G$ and its adjacency matrix $A$, we can get its isomorphic network ${G^{\rm{'}}}$ and corresponding adjacency matrix ${A^{\rm{'}}}$ by renumbered all nodes of $G$. And we can obtain ${A^{\rm{'}}}$ by swapping two rows and corresponding two columns of adjacency matrix $A$ repeatedly. So the main idea of model approximation algorithm can be visualized in Fig.1, where Fig.1(a) is heat map of adjacency matrix $A$ and Fig.1(b) represents ${A^{\rm{'}}}$ which is called block structure or block model denoted as $B$. Then the block ${B_{i,j}}(i < j)$ of block model $B$ in this paper is defined as
\begin{small}
\begin{equation}
{B_{i,j}} = \left[ {\begin{array}{*{20}{c}}
{{b_{i,i}}}&{{b_{i,i + 1}}}& \cdots &{{b_{i,j}}}\\
{{b_{i + 1,i}}}&{{b_{i + 1,i + 1}}}& \cdots &{{b_{i + 1,j}}}\\
 \vdots & \vdots & \ddots & \vdots \\
{{b_{j,i}}}&{{b_{j,i + 1}}}& \cdots &{{b_{j,j}}}
\end{array}} \right]
\label{eq.3}
\end{equation}
\end{small}
\noindent
where ${b_{i,j}} \in B$. Clearly, a block of network $G$ is a sub-matrix of its block model $B$ and the sub-matrix is symmetric around the leading diagonal of $B$. One block in $B$ is a community.

\noindent
\textbf{Definition 4.} d-Strong Block. For an arbitrary threshold value $d \in \left( {0,1} \right)$, the block ${B_{i,j}}$ is a d-Strong Block if

\begin{small}
\begin{equation}
H_{k,l}^{{B_{i,j}}} > d, \forall k,l \in \left( {1,j - i + 1} \right)
\label{eq.3}
\end{equation}
\end{small}
\noindent
In this paper, if all blocks of a block model B satisfy equation (4), then $B$ is a d-Strong block model and denoted as $d-B$ in this paper.

\noindent
\textbf{Definition 5.} d-Weak Block. For an arbitrary threshold value $d \in \left( {0,1} \right)$, the block ${B_{i,j}}$ is a d-Weak Block if

\begin{small}
\begin{equation}
\sum\nolimits_{k,l \in (1,j - i + 1)} {H_{k,l}^{{B_{i,j}}}} > d
\label{eq.5}
\end{equation}
\end{small}

\begin{figure}[htp]
\centering
\includegraphics[width=3.5in]{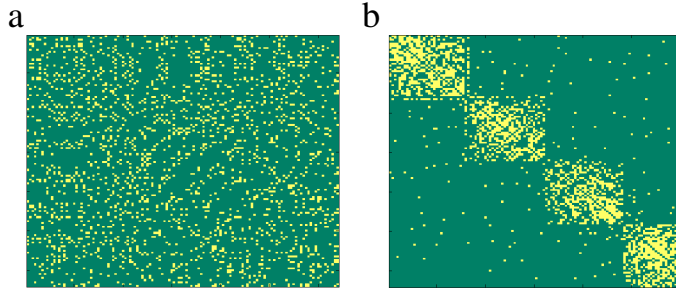}  
\caption{Example network  with non-overlapping communities. (a) Heat map of original adjacency matrix with no blocks. (b) Block model of the network. Clearly, there are four blocks which represent four communities.}
\label{fig.1}
\end{figure}

\subsection{Gravitational Search Algorithm (GSA)}
Inspired by the law of gravity, a novel heuristic optimization algorithm, namely Gravitational Search Algorithm (GSA) was proposed~\cite{39rashedi2009gsa}. It has been proved that GSA is superior to some existing intelligent algorithms including real genetic algorithm, particle swarm optimization and central force optimization. GSA has five main steps which are as follows:

\noindent
\textbf{Step 1}: Initialize Agents. Randomly initialize the positions of $N$ agents as
\begin{small}
\begin{equation}
{X_i} = \left( {x_i^1, \ldots x_i^k, \ldots x_i^h} \right),i = 1,2,\ldots ,N
\label{eq.6}
\end{equation}
\end{small}
where $x_i^k$ is the position of the $i$th agent in the $k$th dimension and $h$ is the dimension of an agent.

\noindent
\textbf{Step 2}: Determine the best and worst fitness of agents. For minimization problems, the best (denoted as $best(t)$) and worst (denoted as $worst(t)$) fitness for all agents at the $t$th iteration are

\begin{small}
\begin{equation}
best(t) = \mathop {\min fi{t_j}\left( t \right)}\limits_{j \in \left\{ {1, \cdots ,N} \right\}}
\label{eq.7}
\end{equation}
\end{small}

\begin{small}
\begin{equation}
worst(t) = \mathop {\max fi{t_j}\left( t \right)}\limits_{j \in \left\{ {1, \cdots ,N} \right\}}
\label{eq.8}
\end{equation}
\end{small}
where $fi{t_j}(t)$ is the fitness value (object function value) of the $j$th agent at the $t$th iteration.

\noindent
\textbf{Step 3}: Calculate and update attribute values of agents.

\noindent%
(1) Gravitational constant $\Psi$. Gravitational constant $\Psi$ is a function of iteration $t$ and defined as

\begin{small}
\begin{equation}
\Psi (t) = {\Psi _0} \times {e^{ - \alpha t/T}}
\label{eq.9}
\end{equation}
\end{small}

where ${\Psi _0}$ is the initial value of gravitational constant, $T$ is the number of iterations and $\alpha$ is a free parameter. According to experiments and analysis in reference~\cite{39rashedi2009gsa}, we set ${\Psi _0} = 100$ and $\alpha =20$ in this paper.

\noindent%
(2) Masses of the agents. At the $t$th iteration, gravitational and inertia masses of agents can be updated by the followings

\begin{small}
\begin{equation}
\left\{ \begin{array}{l}
{M_{a,i}}(t) = {M_{p,i}}(t) = {M_{i,i}}(t) = {M_i}(t)\\
\\
{m_i}\left( t \right) = \frac{{fi{t_i}(t) - worst(t)}}{{best(t) - worst(t)}}\\
\\
{M_i}(t) = {m_i}(t)/(\sum\limits_{j = 1}^N {{m_j}(t)} )
\end{array} \right.
\label{eq.10}
\end{equation}
\end{small}
\noindent%
where ${M_{a,i}}(t)$ is the active gravitational mass and ${M_{p,i}}(t)$ is the passive gravitational mass related to the $i$th agent. ${M_{i,i}}(t)$ represents inertial mass of the $i$th agent.

\noindent%
(3) Acceleration. At the $t$th iteration, the force acting on the $i$th agent from the $j$th agent at the $k$th dimension is

\begin{small}
\begin{equation}
F_{i,j}^k(t) = \Psi (t)\frac{{{M_{p,i}}(t) \times {M_{a,j}}(t)}}{{{R_{i,j}}{{(t)}^{power}} + \varepsilon }}(x_j^k(t) - x_i^k(t))
\label{eq.11}
\end{equation}
\end{small}
\noindent%
where ${R_{i,j}}(t)$ is the Euclidian distance between ${X_i}$ and ${X_j}$, $power$ is a positive integer, $\varepsilon$ is a very small constant. Here we set  $power = 1$, $\varepsilon  = eps$. $eps$ is machine zero in the tool of MATLAB. Then the resultant force of the $i$th agent exerted from other agents at the $k$th dimension is

\begin{small}
\begin{equation}
F_i^k(t) = \sum\limits_{j = 1,j \ne i}^N {ran{d_j} \times F_{i,j}^k(t)}
\label{eq.12}
\end{equation}
\end{small}
\noindent%
where $ran{d_j}$ is a random variable in the interval [0,1]. So the acceleration of the $i$th agent at the $k$th dimension is

\begin{small}
\begin{equation}
a_i^k(t) = \frac{{F_i^k(t)}}{{{M_{i,i}}(t)}}
\label{eq.13}
\end{equation}
\end{small}

\noindent
\textbf{Step 4}: Calculate and update positions and velocities of agents.
\noindent%
After executing the above procedures, the velocities and positions of agents at the $(t+1)$th iteration can be updated by the following equations

\begin{small}
\begin{equation}
\left\{ {\begin{array}{*{20}{c}}
{u_i^k = ran{d_i} \times u_i^k(t) + a_i^k(t)}\\
{x_i^k(t + 1) = x_i^k(t) + u_i^k(t + 1)}
\end{array}} \right.
\label{eq.12}
\end{equation}
\end{small}

\noindent%
where $ran{d_i}$ is a uniform random variable in the interval [0,1], $u_i^k(t)$ and $x_i^k(t)$ are the velocity and position of the $i$th agents in the $k$th dimension at the $t$th iteration, respectively. We set $u_i^k(1) = 0$ for $i = 1,2, \ldots ,N$, $k = 1,2, \ldots ,h$.

\noindent
\textbf{Step 5}: Repeat steps 2 to 4 until $t > T$.

\section{Methods}
\subsection{Recognizing number of communities}
For many community detection algorithms such as partitional clustering methods~\cite{4fortunato2010community}, the number of communities should be given in advance and is often imposed as stopping conditions of algorithms. Identifying gaps between eigenvalues of Laplacian matrix of a network was often applied to estimate the number of communities. However, for networks with indistinct community boundaries and mixed communities, there are not distinct gaps between their Laplacian matrixes’ or Clumpiness matrixes’ eigenvalues. So we creatively utilize $GSA$ algorithm to optimize the proposed $WQ$ value and eventually obtain network block model, then the number of communities (that is strong or weak blocks) can be recognized readily. In order to obtain network block model  $B$, GSA algorithm is employed to search the best sort order of nodes by swapping two rows and corresponding two columns of adjacency matrix $A$  for each circulation. As we all know, $A$ and $B$ have the same eigenvalues and eigenvectors. So we can approximate network adjacency matrix $A$ as $A \approx PB{P^{\rm T}}$, where $P$ corresponds to the top $k$ eigenvectors of $A$ with maximum eigenvalues.

In order to apply GSA to develop the block models of a complex network, there are three problems should be resolved primarily. Firstly, for the problem of initializing agents in GSA, 50 agents are initialized as ${X_i} = \left( {x_i^1, \ldots x_i^k, \ldots x_i^n} \right)$ for $i = 1,2, \ldots ,50$, where $n$ is the number of nodes,  $x_i^k$ is a random integer, $x_i^k \in \{ 1,2, \ldots ,n\} $ and $x_i^k \ne x_i^s$ for $k \ne s$. Secondly, for calculating the fitness value in Step 2 of GSA, the WQ value is utilized as the target function of minimization problem in the optimization model. Thirdly, the best agent we want to obtain is a permutation of integers, but the position of agents may be non-integer after updated in Step 4 of GSA. So the random keys~\cite{41ceberio2012review,42augusto2015pso} is applied to deal with this problem. As an example of the idea of random keys, suppose $x = \left( {\begin{array}{*{20}{c}}{0.5,}&{1.2,}&{0.1,}&{0.3}\end{array}} \right)$ is the updated position of an agent   with four dimensions in Step 4 of GSA. After sorting elements of $x$ , we can get ${x^{''}} = \left( {\begin{array}{*{20}{c}}{0.1,}&{0.3,}&{0.5,}&{1.2}
\end{array}} \right)$. According to the position of every element, the new position of agent $x$ is set as ${x^ * } = \left( {\begin{array}{*{20}{c}}{3,}&{4,}&{1,}&2
\end{array}} \right)$ eventually. For example, the first element of $x$ is 0.5, then it gets to the third position in ${x^{''}}$, so the third dimension is set as 1 in ${x^ * }$. Similarly, the second element 1.2 of $x$ goes to the fourth position in ${x^{''}}$, so the fourth dimension of ${x^ * }$ is set as 2. Other elements of $x$ are processed analogically.

In order to recognize blocks more clearly, a two-steps strategy is proposed including Adding Ones (AO) and Removing Ones (RO). AO is executed by assigning element ${a_{i,j}}$ and all its neighbor cells to one when $H_{i,j}^A$ is greater than a threshold value $d_1 \in \left( {0,1} \right)$. Whereas RO is executed by assigning element ${a_{i,j}}$ and all its neighbor cells to zero when $H_{i,j}^A$ is less than a threshold value $d_2 \in \left( {0,1} \right)$. Then a block model with d-Strong blocks can be obtained eventually.

Moreover, according to the special structure of block model, we calculate the number of 1-elements along each diagonal (line) which parallel to minor diagonal (including minor diagonal). For a $n \times n$ block model B, there are $2n - 2$ diagonals parallel to its minor diagonal and the first one is at the top left of B, the last one at the bottom right. So the value range of X-axis is $\left[ {1,2n - 1} \right]$, and the $(2k - 1)$th line passes through the element ${b_{k,k}}$, we can use this rule to calculate the label of a node according to the label of parallel diagonal.

In this paper, hierarchical communities, overlapping communities and non-overlapping communities are considered in un-weighted and undirected connected networks. Figs.2-4 are block models of three typical networks obtained by GSA algorithm. The RB125 network shown in Fig.2 is a popular synthetic hierarchical and scale-free network~\cite{4fortunato2010community} and it has 25 little communities or 5 large communities. The proposed three-star network shown in Fig.3 has 27 nodes and can be divided into three communities. The networks ${G^{\rm{\# }}}$  with overlapping communities in Fig.4 and non-overlapping communities in Figs.1 are GN benchmark networks, which have 128 nodes and four communities. Clearly, in Fig.2(c), the block model of RB125 has 25 blocks or communities. In Fig.2(b) and Fig.3(b), there are five ''cross'' in block model of RB125 and three ''cross'' in block model of three-star network. As a matter of fact, the “cross” represents community. So we can conclude that RB125 has five communities and the three-star network has three communities. For GN benchmark networks in Fig.4, the block model has four blocks with two pairs of overlapping blocks. So the network has four communities and two overlapping communities in Fig.4.

\begin{figure}[htp]
\centering
\includegraphics[width=4.5in]{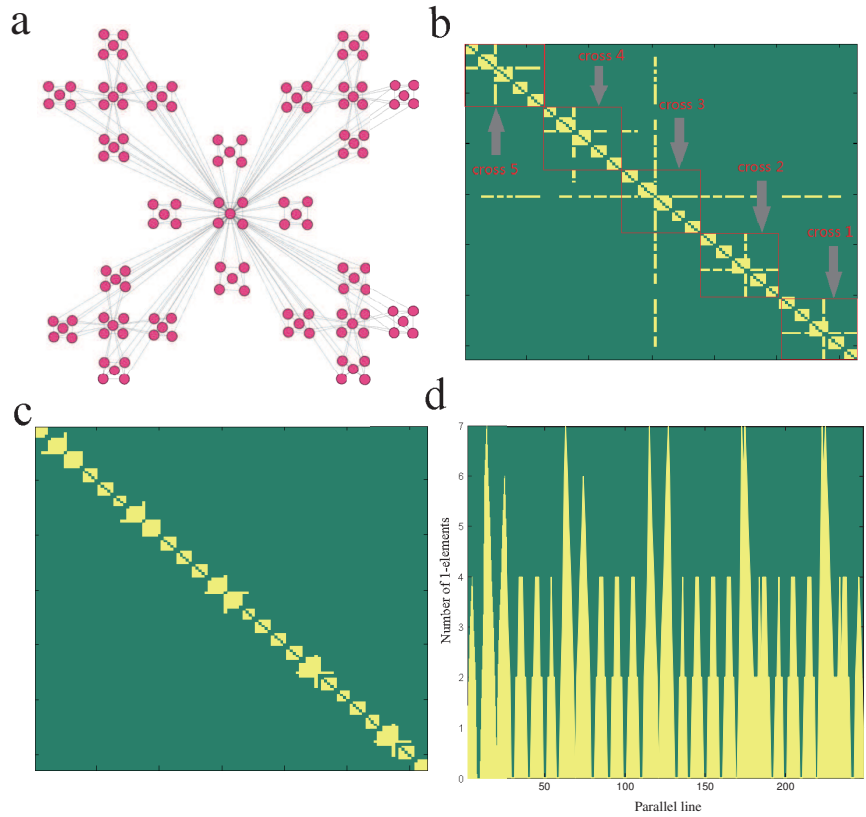}  
\caption{RB125. (a) Visualization of RB125. There are 25 little communities and 5 larger communities. (b) Block model of RB125 obtained by GSA algorithm. Clearly, there are 25 little blocks and 5 ''cross'' structures which represent 5 larger communities. (c) 0.25-B of RB125. This 0.25-Strong block model was obtained by executing AO with $d = 0.7$ and RO with $d = 0.25$. (d) Number of 1-elements along each parallel line. This figure shows relation between the label of parallel lines and number of 1-elements along such parallel lines, where X-axis is the label of parallel lines and Y-axis is the number of 1-elements. The value range of X-axis is $\left[ {1,249} \right]$.}
\label{fig.2}
\end{figure}

\begin{figure}[htp]
\centering
\includegraphics[width=4.5in]{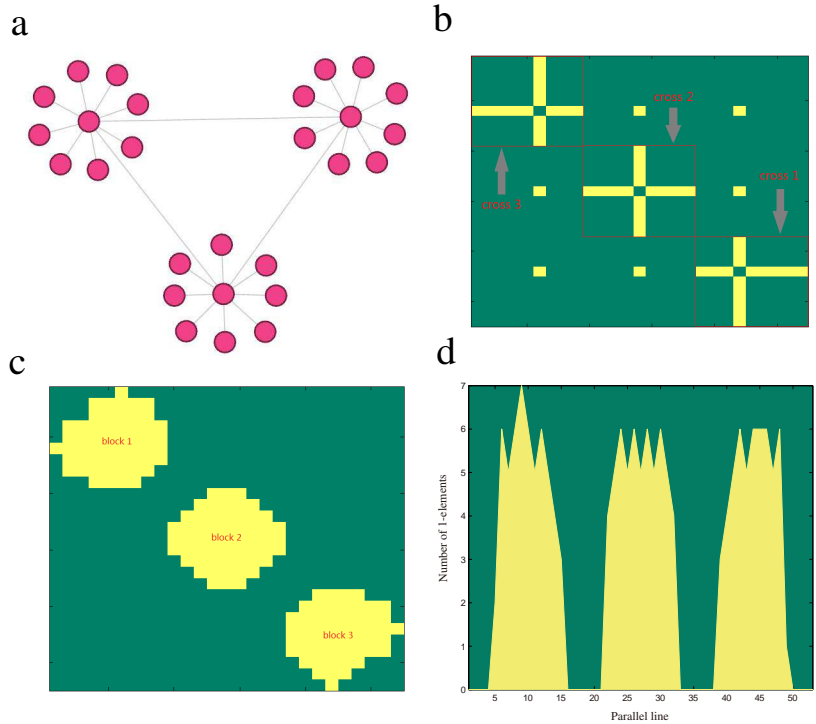}  
\caption{Three-star network. (a) Visualization of three-star network with three communities. (b) Block model of three-star network obtained by GSA algorithm. Clearly, there are three ''cross'' structures which represent three communities. (c) 0.5-B of three-star network. This 0.5-Strong block model was obtained by executing AO with $d = 0.3$ and RO with $d = 0.5$. (d) Number of 1-elements along each parallel line. This figure shows relation between the label of parallel lines and number of 1-elements along such parallel lines, where X-axis is the label of parallel lines and Y-axis is the number of 1-elements. The value range of X-axis is $\left[ {1,53} \right]$. The three peaks mean three communities.}
\label{fig.3}
\end{figure}

\begin{figure}[htp]
\centering
\includegraphics[width=5in]{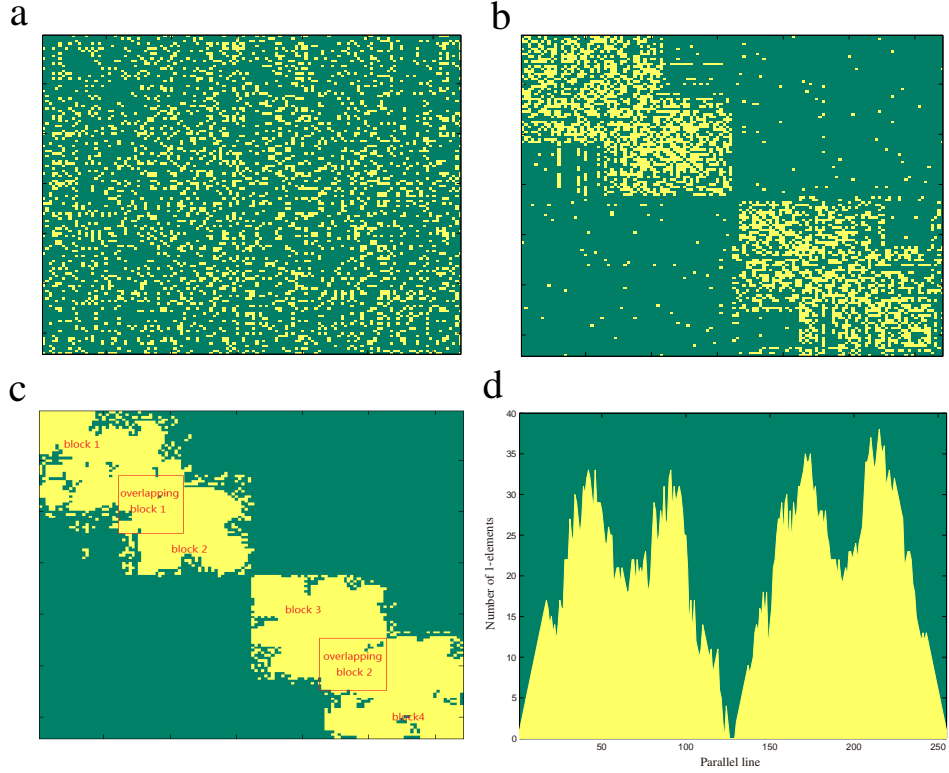}  
\caption{ Example network ${G^{\rm{\# }}}$ with two overlapping communities. (a) Heat map of original adjacency matrix with no blocks. Its WQ value is 0.4274. (b) Block model of ${G^{\rm{\# }}}$  with WQ = 0.4128 obtained by GSA algorithm. Clearly, there are four blocks which represent four communities. Two pairs of blocks are overlapping. (c) 0.2-B of network ${G^{\rm{\# }}}$. This 0.2-Strong block model was obtained by executing AO with $d = 0.5$ and RO with $d = 0.2$. (d) Number of 1-elements along each parallel line. This figure shows relation between the label of parallel lines and number of 1-elements along such parallel lines, where X-axis is the label of parallel lines and Y-axis is the number of 1-elements. The value range of X-axis is $\left[ {1,255} \right]$. There are four small-scale peaks on two large-scale peaks, and it was caused by two overlapping communities. So we can consider this network has two large communities or four little communities with two overlapping communities.}
\label{fig.4}
\end{figure}

\subsection{Detecting community structures}
From the above discussion, we can obtain the block model of a network by the GSA algorithm with the optimal WQ value proposed in this paper. Then the communities can be detected by many existing algorithms such as $k$-means~\cite{4fortunato2010community,34liu2010detecting,35wang2012improved,36jinglu2013finding}, and hierarchical clustering~\cite{4fortunato2010community}. Moreover, these algorithms need to calculate node similarities and even some similarity indices which cost vast time. So we present a novel fuzzy boundary algorithm that can directly process the block model and utilize the proposed fuzzy boundary to detect community structures.

A block model has a corking advantage that its sort order of nodes or rows and columns is in correspondence with the sort order of communities, for example, the first 5 nodes belong to the first community in Fig.2. So the communities can be discovered by detecting boundaries between blocks. We can clearly see that it is hard to observe the distinct boundary between blocks in Fig.2(c), because of the attribution of some nodes could not be decided by our eyeballing. So the set of nodes in a gap between two strong blocks is defined as their fuzzy boundary denoted as ${F_b}$. For a strong block model B of a network, we renumbered all nodes and blocks by their sequence from left to right in the block model. So all nodes from left to right are ${v_1}$, ${v_2}$, ${v_3}$, $ \cdots $, ${v_n}$ and all blocks are ${B_1}$, ${B_2}$, ${B_3}$,$ \cdots $, ${B_k}$ which correspond to communities ${G_1}$,  ${G_2}$,  ${G_3}$, $ \cdots $, ${G_k}$. For $\forall {v_l} \in {F_b}$, $\exists {B_i},{B_{i + 1}}$, and for $\forall {v_p} \in {B_i}$ and $\forall {v_q} \in {B_{i{\rm{ + }}1}}$ we can obtain $p \le l \le q$.

After obtaining all ${F_b}$, the rest blocks which are separated by ${F_b}$ compose initial community structures $\left\{ {{G_1},{G_2}, \ldots ,{G_k}} \right\}$, where $k$ is the number of blocks. In order to distribute nodes in ${F_b}$, the similarity between a node and blocks (communities) is defined as the number of links between them. Moreover, the nodes in ${F_b}$ belong to the adjacent block which has the greatest similarity with them, and the nodes with identical similarity are overlapping nodes. After each circulation of distributing nodes in ${F_b}$, all similarities of remaining nodes should be updated until ${F_b}$ is empty or all nodes in ${F_b}$ are overlapping nodes. We call this strategy of detecting communities as fuzzy boundary algorithm and it can be described as follows:

\noindent
\textbf{Input}: Network $G = (V,E)$

\noindent
\textbf{Output}: Communities $\left\{ {{G_1},{G_2}, \ldots ,{G_k}} \right\}$

\noindent
\textbf{Step 1}: Use GSA algorithm to optimize the WQ value of adjacency matrix $A$ to get block model $B$.

\noindent
\textbf{Step 2}: Obtain $d$-Strong block model $d$-B by applying AO and RO to $B$.

\noindent
\textbf{Step 3}: Detect ${F_b}$ and initial community structures $\left\{ {{G_1},{G_2}, \ldots ,{G_k}} \right\}$ of $d$-B.

\noindent
\textbf{Step 4}: Calculate similarities between nodes in ${F_b}$ and $\left\{ {{G_1},{G_2}, \ldots ,{G_k}} \right\}$, $i = 1,2, \ldots ,k$.

\noindent
\textbf{Step 5}: Distribute nodes in ${F_b}$ to the adjacent communities and update $\left\{ {{G_1},{G_2}, \ldots ,{G_k}} \right\}$ and ${F_b}$.

\noindent
\textbf{Step 6}: Repeat steps from Step 4 to Step 6 until ${F_b} = \emptyset $.

\noindent
\textbf{Step 7}: Calculate similarities between nodes and communities, adjust misallocated nodes.

\noindent
\textbf{Step 8}: Output communities $\left\{ {{G_1},{G_2}, \ldots ,{G_k}} \right\}$.

\subsection{Experimental results}
The fuzzy boundary algorithm is applied to two representative networks including Football team network~\cite{43palla2005uncovering} and pol-blog network~\cite{6adamic2005political}, and experimental results are compared with some existing methods by modularity Q and NMI. Experimental results are shown in Figs.5-7. The Football team network contains 115 nodes and 613 edges, nodes are college football teams and edges are competition schedule of teams, and it has 12 communities originally. The pol-blog network, with 1490 nodes and 19090 links, represents a set of web-blogs of US politics. This network primitively has two communities including conservative and liberal, and its maximal connected network with 1222 nodes and 16714 links is discussed in this paper.

On Football team network, the block model B has 11 blocks shown in Fig.5, then 11 communities are discovered with WQ=0.1420 which is less than WQ= 0.4271 of original adjacency matrix A. After executing AO with d=0.9 and RO with d=0.6, block model 0.6-B is obtained, and its corresponding ${F_b} = \{
{v_{30}},{v_{35}},{v_{27}},{v_{37}},{v_{40}},{v_7},{v_{65}},{v_{107}},{v_{90}},{v_2},{v_{43}},{v_{81}},{v_{83}},{v_{64}},{v_{63}},{v_{77}},{v_{60}},{v_{114}},
{v_{88}},$

\noindent
${v_{59}},{v_{45}},{v_{58}},{v_{92}},{v_{49}},{v_{98}},{v_4},{v_{103}},{v_{12}},{v_{25}},{v_{29}},{v_{70}},{v_{54}},{v_{68}},{v_{91}},{v_{51}},{v_1},{v_{24}},{v_8}\} $, 11 communities are detected and shown in Fig.6. The optimal modularity Q=0.5621 and NMI=0.9043. Modularity of initial communities is Q = 0.5540. This network is also discussed in~\cite{14newman2004fast}, and its optimal modularity is Q = 0.546 and NMI=0.6838 with 6 communities. In~\cite{44girvan2002community}，their optimal modularity is Q=0.6005 and NMI=0.9095 with 11 communities.

From Fig.6, the proposed fuzzy boundary algorithm can successfully detect communities ${G_1}$, ${G_2}$, ${G_3}$, ${G_4}$, ${G_5}$, ${G_6}$ and ${G_9}$ of original communities. Community ${G_7}$ is divided into two communities, and nodes of communities ${G_{11}}$ and ${G_{12}}$ are distributed to other communities. In football team network, the initial 12 communities are obtained by football teams from different geographical location, and their links represent schedule of competition between teams. So in initial community structures, several nodes such as ${v_{29}}$, ${v_{37}}$ and ${v_{111}}$ may have more intercommunity links than intra-community links. Then experimental results may have some distinctions compared with initial community structure. In general, the fuzzy boundary algorithm is feasible.

\begin{figure}[htp]
\centering
\includegraphics[width=4.5in]{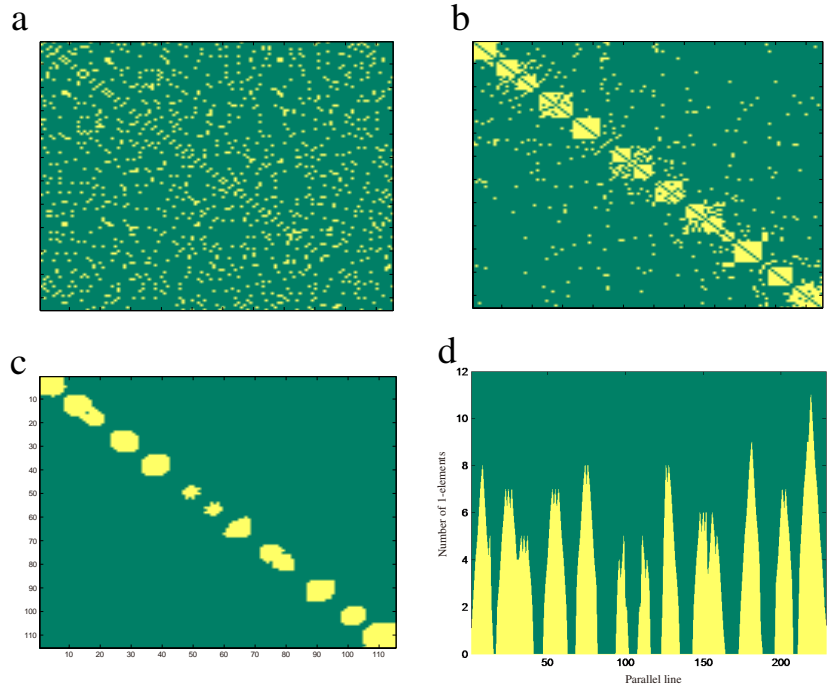}  
\caption{Football team network. (a) Visualization of original 12 communities of football team network. Different colors represent different communities. (b) Heat map of original adjacency matrix with no blocks. Its WQ value is 0.4271. (c) Block model of football team network obtained by GSA algorithm. Its WQ value is 0.1420 which is less than WQ in (b). (d) 0.7-B of football team network. This 0.7-Strong block model was obtained by executing AO with $d = 0.7$ and RO with $d = 0.7$. (e) Number of 1-elements along each parallel line. This figure shows relation between the label of parallel lines and number of 1-elements along \protect\\ such parallel lines, where X-axis is the label of parallel lines and Y-axis is the number of 1-elements. The value range of X-axis is $\left[ {1,229} \right]$. There are 11 non-overlapping peaks and it means 11 non-overlapping communities. According to gaps between peaks, ${F_b}$ can be obtained.}
\label{fig.5}
\end{figure}

\begin{figure}[htp]
\centering
\includegraphics[width=4in]{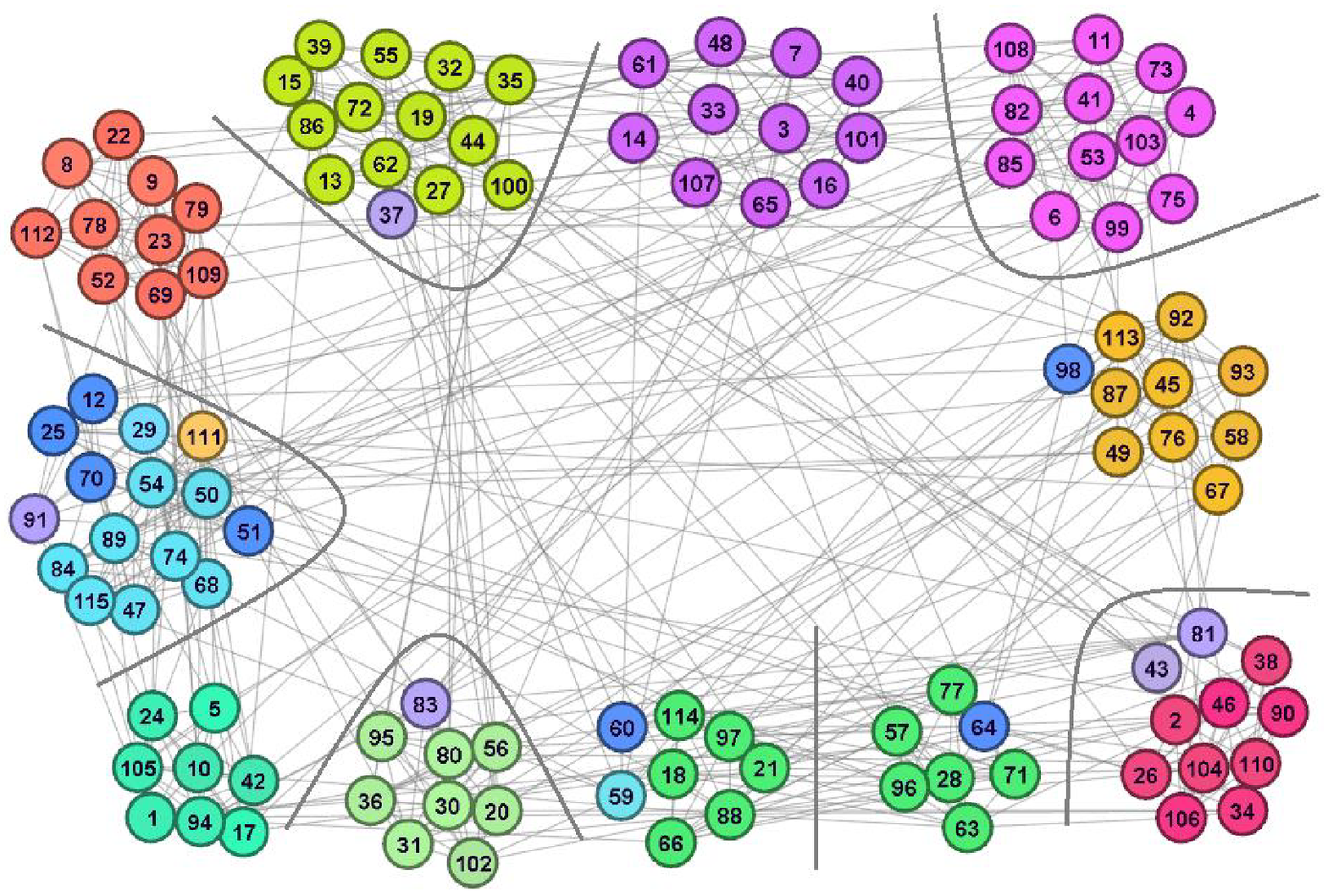}  
\caption{ Community structure of football team network detected by the proposed fuzzy boundary algorithm. 11 communities are detected with WQ=0.1420, modularity Q=0.5514 and NMI=0.8827.Nodes of the same color originally belong to the same community.}
\label{fig.6}
\end{figure}

Two communities are detected from pol-blog network shown in Fig.7, and the optimal WQ value is 0.5803 which is less than 0.6804 of original adjacency matrix A. After executing AO with d=0.2 and RO with d=0.9, 0.9-Strong block model 0.9-B is obtained. Because there are two communities in pol-blog network, the labels of nodes in ${F_b}$ correspond to the labels of parallel lines from 1292th to 1450th, then ${F_b}$ can be obtained by transforming parallel line label to node label. Eventually, we get two communities with modularity Q=0.4052. This network was also discussed in~\cite{1wang2015novel} and~\cite{45raghavan2007near}, their optimal modularity are Q=0.424236 and Q=0.4176, respectively. Actually, it can be found from Fig.7(a) that two blocks exist in the heat map of original adjacency matrix, so experimental results of the proposed fuzzy boundary algorithm are reasonable.

\begin{figure}[htp]
\centering
\includegraphics[width=5in]{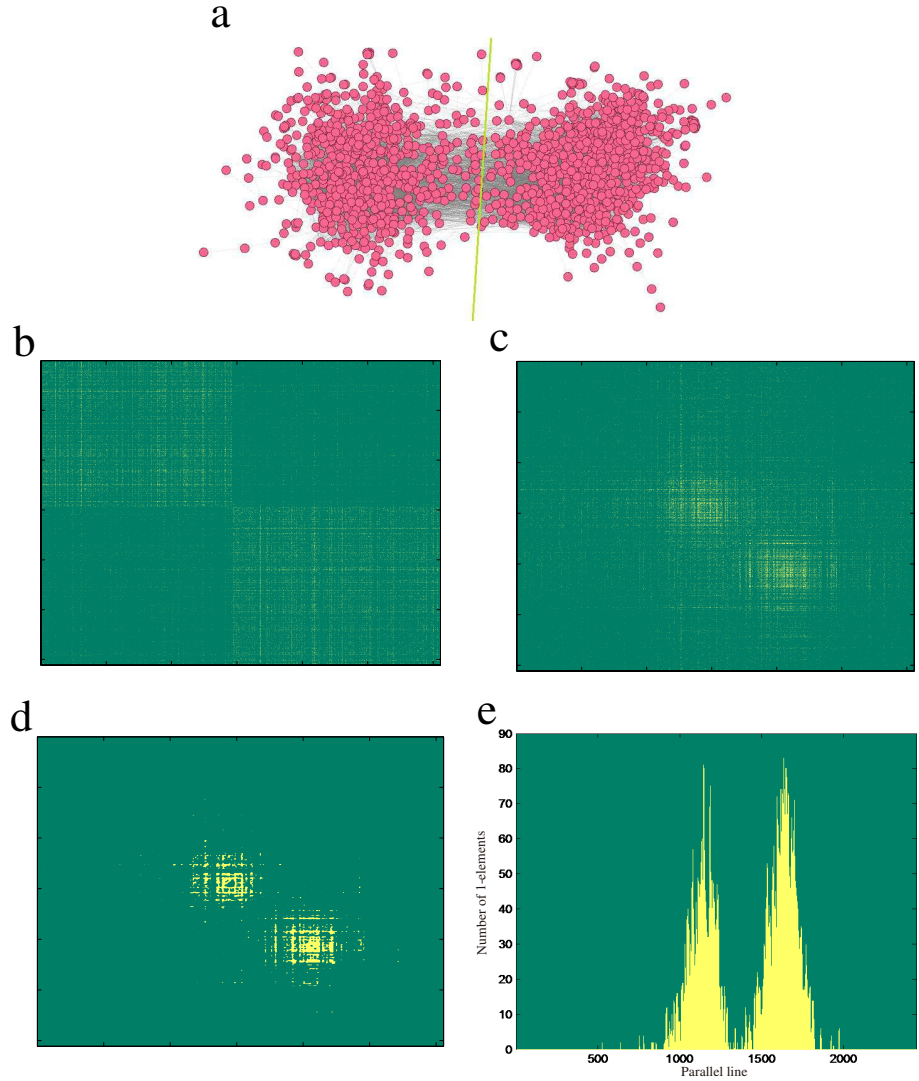}  
\caption{ Pol-blog network. (a) Visualization of original two communities of pol-blog network. (b) Heat map of original adjacency matrix with two blocks. Its WQ value is 0.6804. (c) Block model of pol-blog network obtained by GSA algorithm. It has two blocks and its WQ value is 0.5803 which is less than WQ in (b). (d) 0.9-B of football team network. This 0.9-Strong block model was obtained by executing AO with d=0.2 and RO with  d=0.9. (e) Number of 1-elements along each parallel line. This figure shows relation between the label of parallel lines and number of 1-elements along such parallel lines, where X-axis is the label of parallel lines and Y-axis is the number of 1-elements. The value range of X-axis is  $\left[ {1,2443} \right]$.}
\label{fig.7}
\end{figure}

\section{Conclusion}
In this paper, in order to measure the quality of block model, we first define an objective function WQ value. For obtaining block model B of a network, GSA algorithm is applied to optimize WQ with the help of random keys. After executing AO and RO on block model B, the number of communities of a network can be recognized distinctly. Furthermore, based on the advantage of block model that its sort order of nodes or rows and columns is in correspondence with sort order of communities, an algorithm for detecting community structures called Fuzzy Boundary is proposed and successfully applied to some representative networks. Finally, experimental results demonstrate the feasibility of our algorithm. There are some rooms need to improve in the future. Firstly, the threshold value d in AO and RO needs an in-depth study, for example, AO may merge some block together with too small threshold and RO may break a block into more with too large threshold. Secondly, for some large-scale network, GSA has high time-complexity and the problem of trapping in local optimum. So the experimental results in this paper may not be optimal division but feasible. Thirdly, if the number of nodes in ${F_b}$ is overmuch, cohesiveness of existing communities might be weakened. What’s more, other similarity indices between nodes and communities can be applied to distribute nodes in ${F_b}$. Finally, we hope the algorithm presented here can be expanded to other types of networks, such as weighted network, directed network, bipartite network and dynamic network. We hope such improvements and more applications of the proposed algorithm in the future.


\bibliographystyle{unsrt}
\bibliography{reference}

\end{document}